\documentclass[twocolumn,showpacs,preprintnumbers,amsmath,amssymb]{revtex4}
\usepackage{graphicx}
\begin{document}
\title{Comparison of mid-velocity fragment formation with projectile-like decay}

\author{S. Hudan}
\author{R. Alfaro}
\author{B. Davin}
\author{Y. Larochelle}
\author{H. Xu}
\altaffiliation[]{Present address: Institute of Modern Physics, CAS, Lanzhou, China.} 
\author{L. Beaulieu}
\altaffiliation[] {Present address: Universit\'{e} Laval, Quebec, Canada.}
\author{T. Lefort}
\altaffiliation[] {Present address: Universit\'{e} de Caen, Caen, France.} 
\author{R. Yanez}
\altaffiliation[] {Present address: Department of Nuclear Physics, 
The Australian National University, 
Canberra, Australia.} 
\author{R.T. de Souza} 
\affiliation{
Department of Chemistry and Indiana University Cyclotron Facility \\ 
Indiana University, Bloomington, IN 47405} 

\author{R.J. Charity}
\author{L.G. Sobotka}
\affiliation{
Department of Chemistry, Washington University, St. Louis, MO 63130}

\author{T.X. Liu}
\author{X.D. Liu}
\author{W.G. Lynch}
\author{R. Shomin}
\author{W.P. Tan}
\author{M.B. Tsang} 
\author{A. Vander Molen} 
\author{A. Wagner}\altaffiliation[] {Present address: Institute of Nuclear and Hadron Physics, Dresden, Germany. } 
\author{H.F. Xi}
\affiliation{
National Superconducting Cyclotron Laboratory and Department of
Physics and Astronomy \\ 
Michigan State University, East Lansing, MI 48824}

\date{\today}

\begin{abstract}
The characteristics of intermediate mass fragments (IMFs: 3$\leq$Z$\leq$20) 
produced in  mid-peripheral and central collisions are compared. 
We compare IMFs detected at mid-velocity with those evaporated
from the excited 
projectile-like fragment (PLF$^*$). 
On average,
the IMFs produced at mid-velocity are 
larger in atomic number, exhibit broader transverse velocity 
distributions, and 
are more neutron-rich as
compared to IMFs evaporated from the PLF$^*$. 
In contrast, comparison of mid-velocity fragments 
associated with mid-peripheral and central collisions
reveals that their 
characteristics are remarkably similar despite the difference 
in impact parameter. The characteristics of mid-velocity fragments 
are consistent with 
low-density formation of the fragments.
Neutron deficient isotopes of even Z elements manifest higher kinetic 
energies than heavier isotopes of the same element
for both PLF$^*$ and mid-velocity emission. 
This result may be 
due to the decay of long-lived excited states 
in the field of the emitting system.

\end{abstract}

\maketitle

\section{Introduction}

Cluster emission from nuclear matter can arise from a wide range of
phenomena, such as statistical evaporation from normal density 
nuclear matter at modest excitation 
\cite{Sobotka83} or the multi-fragmentation of low-density nuclear matter at 
high excitation induced by GeV hadronic projectiles \cite{Beaulieu00}.
Collision of two
heavy-ions at intermediate energies (25 MeV $\leq$E/A$\leq$100 MeV) 
also results in 
copious intermediate mass fragment (IMF : 3$\leq$Z$\leq$20)  
production \cite{deSouza91,Bowman91}. 
Considerable attention has been focused on understanding 
the conditions governing the maximum fragment yield \cite{Ogilvie91,Peaslee94}
and the characteristics of the fragments produced \cite{Milkau91,Kim91}. 
In peripheral collisions of two intermediate-energy (20$\le$E/A$\le$100 MeV) 
heavy nuclei (A$\sim$100) a dissipative binary collision 
occurs resulting in the formation of an excited 
projectile-like fragment (PLF$^*$) and  target-like fragment (TLF$^*$). 
The dominant IMF yield in such collisions is observed at velocities 
intermediate
between the de-excited PLF$^*$ and TLF$^*$, 
and is not-attributable to the standard statistical
decay of either of the two reaction partners \cite{Toke95,Plagnol00}. 
The IMFs in this kinematical region are referred to as
mid-velocity IMFs. 
For more central collisions,
the distinctive binary nature of the collision is no longer apparent, 
nevertheless 
most of the IMF emission occurs in the 
same kinematical region as in more peripheral collisions. 
While for more peripheral
collisions, the dynamical nature of mid-velocity fragments has been shown
\cite{Bocage00,Davin02,Piantelli02,Colin03},  
in the case of central collisions, 
statistical approaches are generally used to understand the fragment production
\cite{Marie97,Milazzo02a,Xu00}.

On general grounds, the size, composition, and kinetic energies of the 
observed clusters, apart from their yield, can be related to
the composition and excitation of the disintegrating system. For example,
the composition of fragments, namely their neutron-to-proton ratio, may provide
information on the N/Z of the disintegrating system \cite{Colonna02,Sil04}.
Several experiments have established 
the neutron enrichment of IMFs and light clusters (Z$\le$2) 
at mid-velocity
\cite{Dempsey96,Xu00,Larochelle00,Milazzo02a}.
The observation of neutron-rich fragments in this kinematic region  
has been interpreted as the N/Z fractionation
of hot nuclear material into a neutron-rich gas and a proton-rich 
liquid \cite{Xu00}. As with any claim of ``enrichment'' of a quantity, it is 
necessary to establish the appropriate reference with respect to which the 
enrichment occurs. We propose that the most appropriate reference for 
investigating possible enrichment of mid-rapidity fragments is the N/Z of 
the emitted fragments from near normal density nuclear matter. 

In this work, we examine
the fragment characteristics
largely independent of the probability of their formation.
We show that, at mid-velocity, the fragment characteristics 
manifest significant differences as 
compared to those evaporated from near normal
density nuclear matter.
In contrast, similar fragment characteristics at mid-velocity are 
observed both for mid-peripheral collisions and central 
collisions. 
In contrast to previous results \cite{Milazzo02a}, the size of 
the emitting system is shown to {\it not} be the determining 
factor in the composition
of the emitted fragments.

    To investigate the factors influencing fragment composition, we measured 
IMF and light-charged-particle (LCP:1$\leq$Z$\leq$2) emission in the reaction
$^{114}$Cd + $^{92}$Mo at E/A=50 MeV. We examine 
mid-peripheral collisions in which
the survival of a well defined  
projectile-like fragment occurs. Emission from the
PLF$^*$  (which presumably is at near normal density) provides
a suitable reference for understanding mid-velocity IMF emission in the same collisions. 
We subsequently  
compare mid-velocity IMFs associated with
mid-peripheral collisions and those associated with central collisions.

\section{Experimental Setup}

\begin{figure}
\vspace*{2.5in} 
\includegraphics{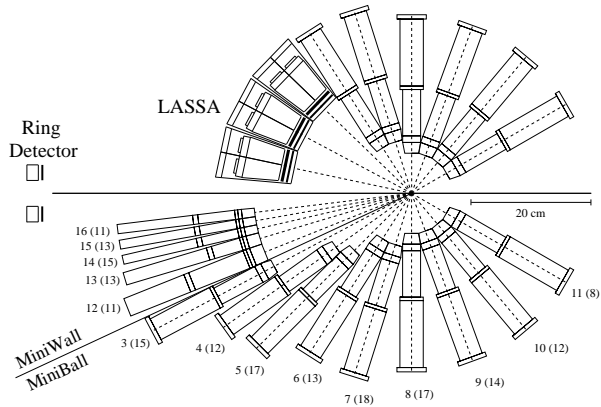}
\caption[]
{Experimental setup used to study the reaction $^{114}$Cd+$^{92}$Mo 
at E/A=50 MeV. The number of detectors in each azimuthal ring of the Miniball/Miniwall array 
is indicated in parentheses.} 
\label{fig:setup}
\end{figure}

\begin{figure}
\vspace*{3.0in} 
\includegraphics{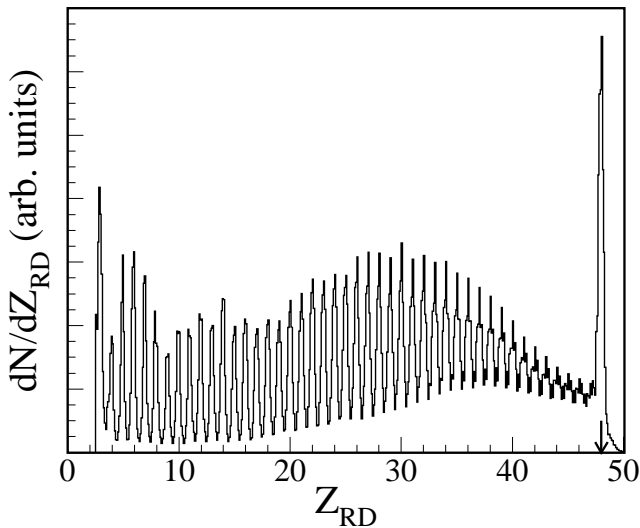}
\caption[]
{Element distribution measured by the ring detector for the angular range
2.1$^{\circ}$$\leq$$\theta^{lab}$$\leq$4.2$^{\circ}$. The arrow indicates
the atomic number of the beam.} 
\label{fig:rcreso}
\end{figure}

\begin{figure}
\vspace*{3.5in} 
\includegraphics{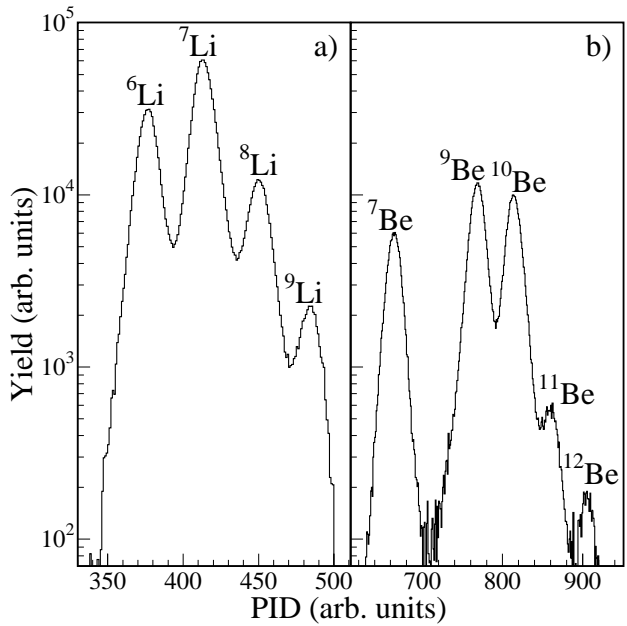}
\caption[]
{Isotopic resolution achieved in LASSA for isotopes of Li and Be. 
The spectra have been summed over all nine LASSA telescopes.} 
\label{fig:LASSAresol}
\end{figure}

Charged-particles produced in the reaction $^{114}$Cd+$^{92}$Mo at E/A=50 MeV 
were detected in the 
exclusive 4$\pi$ setup depicted in Fig.~\ref{fig:setup}. 
Peripheral collisions were selected by the detection of 
forward-moving projectile-like fragments (PLFs).
These PLFs were detected in the angular range 
2.1$^{\circ}$$\leq$$\theta^{lab}$$\leq$4.2$^{\circ}$ and were 
identified in an annular Si(IP)/CsI(Tl)/PD ring detector (RD) by the 
$\Delta$E-E technique. This telescope provided elemental identification  
with better than unit Z resolution for Z$\leq$48, as shown in 
Fig.~\ref{fig:rcreso}. 
The peak at Z=48 corresponds to quasi-elastically scattered projectile nuclei 
associated with the most peripheral collisions.
The silicon $\Delta$E element of this telescope 
was segmented into 16 concentric rings on its junction side and 16 pie-shaped 
sectors on its ohmic surface. The ring segmentation
provided a good measurement of the polar angle of the PLF,
typically $\Delta\theta^{lab}$$<$0.2$^{\circ}$,
while the pie-shaped sectors allowed a measure of 
the azimuthal angle \cite{Davin02}. 
Careful calibration of the CsI(Tl) crystals with 70 fragmentation beams 
allowed determination of the light response of the CsI(Tl) crystals 
resulting in a typical energy resolution of 3$\%$. 
From the measured Z, angle, and energy, the velocity of the PLF was calculated
by assigning the A for a given Z consistent with systematics \cite{Summerer90}
adjusted near Z$_{beam}$ to correspond to the N/Z of the projectile 
\cite{davin_thesis}.
Intermediate mass fragments with Z$\le$9 and
light-charged-particles were
isotopically identified in the angular range 
7$^{\circ}$$\leq$$\theta^{lab}$$\leq$58$^{\circ}$ with the high 
resolution silicon-strip array LASSA \cite{Davin01,Wagner01}. 
Each of the nine 
telescopes in this array consisted of a stack of three elements, two 
ion-implanted, passivated silicon strip detectors (Si(IP)) backed by a
2 x 2 arrangement of
CsI(Tl) crystals each with photo-diode readout. The second silicon of each telescope 
was segmented into 16 vertical strips and 16 
horizontal strips, resulting in good angular resolution 
($\Delta$$\theta^{lab}$$\approx$0.43$^\circ$). 
The nine LASSA telescopes were arranged in a 3 x 3 array, the center 
of which was located at a polar angle 
$\theta^{lab}$=32$^\circ$ with respect to the beam axis.
The energy threshold of LASSA is 2 and 4 MeV/A for 
$\alpha$ particles and carbon fragments, respectively. A typical example of
the isotopic resolution
achieved by LASSA is shown in Fig.~\ref{fig:LASSAresol}. 
Isotopes of Li and Be are clearly resolved with an energy resolution 
of $\approx$2-5$\%$.
In order to augment the limited kinematical coverage of LASSA and the RD, 
the low-threshold Miniball/Miniwall array \cite{Miniball} was used to 
identify charged-particles emitted in the range 
5$^{\circ}$$\leq$$\theta^{lab}$$\leq$168$^{\circ}$.
Using pulse-shape discrimination, particles detected in the 
Miniball/Miniwall array were isotopically identified for Z$\le$2. These 
particles 
were used to select the impact parameter of the collision  
and to globally characterize the selected events.

\section{General Reaction Characteristics and Event selection}

\begin{figure}
\vspace*{3.5in} 
\includegraphics{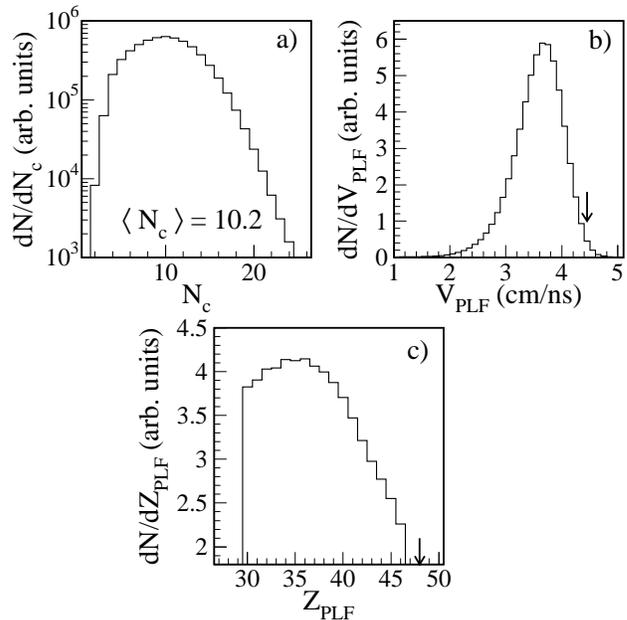}
\caption[]
{Panel a) Charged-particle multiplicity distribution associated with the detection of
a PLF with 30$\le$Z$\le$46 in the RD; Panel b) Velocity
distribution for 
30$\le$Z$\le$46 detected in RD; Panel c) Atomic number distribution 
of fragments detected 
in the RD.} 
\label{fig:global_peri}
\end{figure}

We begin by examining mid-peripheral (MP) events 
distinguished by 
the survival of a projectile-like fragment at forward angles. In order 
to examine these peripheral collisions we have
selected events in which a heavy PLF with 30$\le$Z$\le$46 is 
detected in the RD (2.1$^{\circ}$$\leq$$\theta^{lab}$$\leq$4.2$^{\circ}$) 
coincident with at least three charged-particles 
in the Miniball/Miniwall array. This latter charged-particle requirement 
suppresses the most peripheral collisions and results in the associated
multiplicity distribution shown in Fig.~\ref{fig:global_peri}a.
These MP collisions are characterized by
an average total charged-particle 
multiplicity, $\langle$N$_C$$\rangle$, of 10.2, with a 
second moment ($\mu_2$) of 3.6. 
Based on the charged-particle multiplicity distribution
\cite{Cavata90}, we estimate the impact-parameter ratio 
b/b$_{max}$ $\approx$ 0.7 where b$_{max}$ 
represents the interaction for which at least three charged-particles 
are detected in the Miniball/Miniwall array.
The center-of-mass velocity distribution of the PLF detected in the RD is
shown in Fig.~\ref{fig:global_peri}b
with the beam velocity indicated by an arrow for reference. One observes that this
distribution is a skewed gaussian with a tail 
toward lower velocities. 
The most probable value of this velocity distribution is
3.7 cm/ns ($\langle$V$_{PLF}$$\rangle$=3.57 cm/ns), indicating 
an average velocity damping of 0.88 cm/ns from the beam velocity. 
The atomic number distribution of PLFs associated with these collisions is
displayed in Fig.~\ref{fig:global_peri}c. 
The most probable value 
of Z$_{PLF}$ is Z=35 ($\langle$Z$_{PLF}$$\rangle$=37) as compared 
to Z$_{beam}$=48, indicated by the arrow. It should be realized that this 
most probable (or average) Z$_{PLF}$ corresponds to the PLF following the 
de-excitation of the primary excited projectile-like fragment (PLF$^*$).

\begin{figure}
\vspace*{4.5in} 
\includegraphics{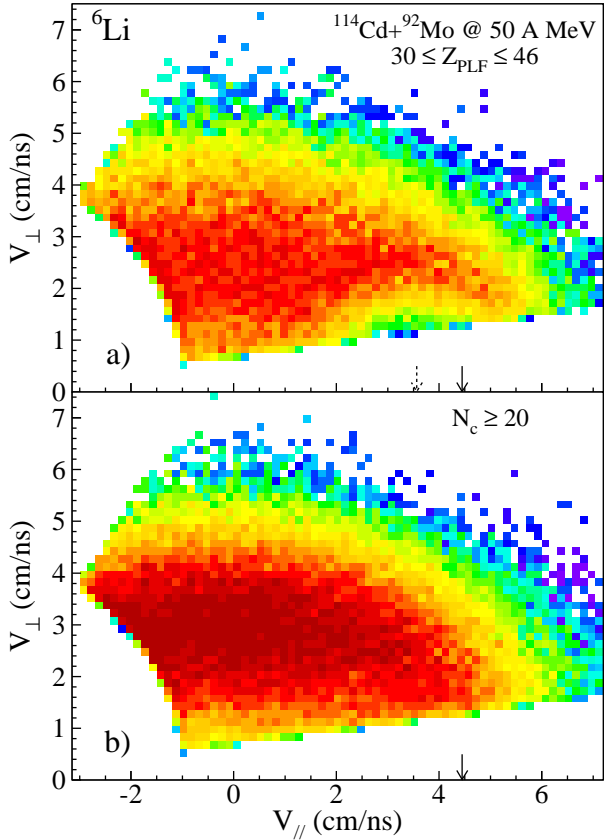}
\caption[]
{(Color online) Invariant cross-section plots of the emission of $^6$Li fragments in the
center-of-mass frame. Panel a)
associated with 30$\leq$Z$_{PLF}$$\leq$46. Panel b)
associated with N$_{C}$$\ge$20.
The dashed arrow indicates 
$\langle$V$_{\parallel}^{PLF}$$\rangle$ 
while the solid arrow indicates
V$_{BEAM}$. The differential yield is presented on a logarithmic scale.} 
\label{fig:gali}
\end{figure}

The general characteristic of this de-excitation of the PLF$^*$
is shown in Fig.~\ref{fig:gali}a. 
Clearly evident in Fig.~\ref{fig:gali}a is  a circular
ridge of yield 
centered at V$_{\parallel}$ 
$\approx$ 3.5 cm/ns in the center-of-mass frame. 
This ridge can be understood as 
emission of $^6$Li fragments from the PLF$^*$,
following the interaction phase of the reaction.
This distinctive emission pattern indicates that for the 
collisions selected, a binary reaction has 
occured \cite{Lott92}. 
By utilizing the measured multiplicities, kinetic energy spectra, and angular 
distributions of particles detected in coincidence with the PLF, we have
reconstructed (under the assumption of isotropic emission) 
the average atomic number of the PLF$^*$, $\langle$Z$_{PLF*}$$\rangle$ 
and its excitation
\cite{Yanez03}. For the collisions studied, we have determined 
that $\langle$Z$_{PLF*}$$\rangle$$\approx$41.
Also clearly evident in Fig.~\ref{fig:gali}a, and well established by 
earlier work \cite{Toke95,Plagnol00}, is that considerable fragment emission occurs
at mid-velocity -- emission {\it not} originating from the isotropic 
statistical decay of the PLF$^*$ or TLF$^*$ reaction partners
\cite{Bocage00}. For the remainder of this work we define mid-velocity fragments as
those with 0$\le$V$_{\parallel}$$\le$1cm/ns in the center-of-mass frame.

\begin{figure}
\vspace*{3.5in} 
\includegraphics{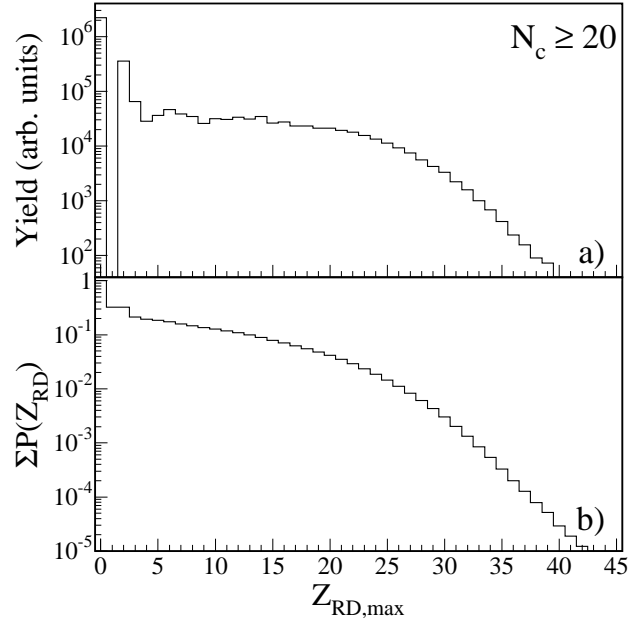}
\caption[]
{Panel a) Z distribution of the largest fragment detected in the RD associated 
with events for which N$_c$$\ge$20.
Panel b) cumulative yield distribution of the largest fragment in the RD associated with events for which N$_c$$\ge$20.
} 
\label{fig:zplf_central}
\end{figure}

For central collisions (N$_C$$\ge$20 ; $\langle$N$_C$$\rangle$=21.8;
$\mu_2$=1.89) the ridge centered near the projectile velocity is no longer 
observed, as shown in Fig.~\ref{fig:gali}b. 
The observation that a clear Coulomb circle does not exist 
has traditionally been interpreted as
evidence that the collision is no longer a binary process.
However, the observation of a large charged particle multiplicity together
with the absence of a Coulomb circle does not preclude the existence of a 
dissipative binary process. Rather, these observations can be reconciled 
with the rapid de-excitation of the PLF$^*$ and TLF$^*$ on a timescale
commensurate with their re-separation.
In contrast to the well defined Coulomb circle of 
Fig.~\ref{fig:gali}a, the emission pattern for central collisions is 
broad and featureless with
substantial emission near the center-of-mass velocity. In these collisions we deduce 
from the charged-particle multiplicity that
b/b$_{max}$= 0.26, from which 
we estimate Z$_{source}$ $\approx$ 72 \cite{Xu02}. 
The charged-particle multiplicity has often been used in this 
manner to select central collisions.
Examination of the 
Z distribution of the largest measured particle in the RD associated with 
these events, however, is quite 
revealing. As evident in Fig.~\ref{fig:zplf_central}a, 
while the largest probabilities are associated with either the 
detection of no fragment 
(Z=0) or a helium in the RD, 
there is a 
significant probability of detecting a large fragment with Z $\ge$10 in the RD.
(It should be noted that the triggering threshold in the RD was set to not 
trigger on hydrogen nuclei resulting in a measured yield of zero for Z=1 in 
Fig.~\ref{fig:zplf_central}a.) 
This detection of a large fragment (Z$\ge$10) occurs even though a large
charged-particle multiplicity has been required. This result 
is consistent with the physical picture of a large overlap of 
projectile and target nuclei which still results in 
a binary exit channel with survival of a projectile-like 
and target-like fragment. The cumulative yield associated with such events,
$\Sigma$P(Z$_{RD}$), is 
shown in Fig.~\ref{fig:zplf_central}b where 
\begin{equation}
\Sigma P(Z_{RD})=\int\limits_{Z=45}^{Z_i} \frac{dN}{dZ_{RD,max}} dZ
\end{equation}
Evident in Fig.~\ref{fig:zplf_central}b is the result that
$\Sigma$P(Z$_{RD}$) $\approx$ 0.1 for Z$_{RD,max}$=10.
This result reveals that $\approx$ 10 $\%$ of the central collisions selected 
simply by the requirement that N$_C$$\ge$20 are in fact associated with binary  
collisions. Due to the limited angular acceptance of the RD, 
this``contamination'' of true central collisions might be somewhat higher.

\begin{figure}
\vspace*{3.5in} 
\includegraphics{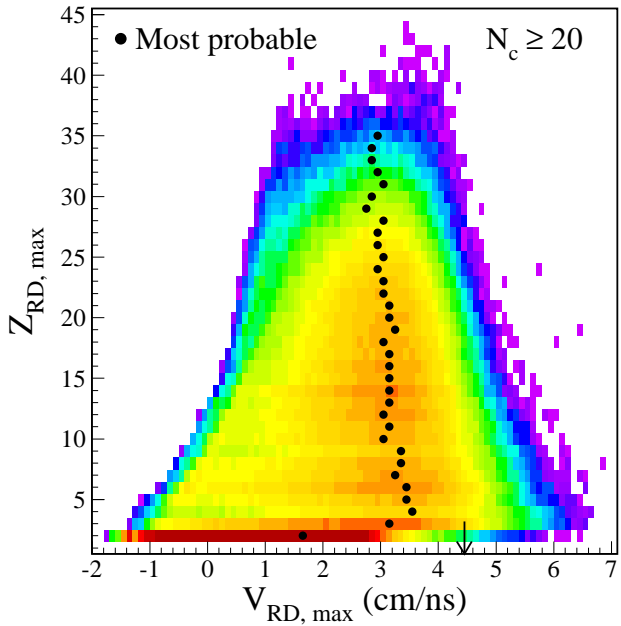}
\caption[]
{(Color online) Correlation between atomic number and velocity (in the COM) 
of the largest fragment 
detected in the RD. The most probable velocity for each Z is indicated by 
a filled circle while the beam velocity is indicated by the arrow.} 
\label{fig:zvrc_max_central}
\end{figure}

Further evidence that this ``contamination'' is associated with a 
binary exit channel and not simply an anisotropic emission pattern is provided 
in Fig.~\ref{fig:zvrc_max_central}. In this figure, we examine the 
correlation between the atomic number and the velocity of these fragments 
detected in the RD. With the exception of the lightest fragments (Z$\le$3), 
the fragments detected in the RD have a most probable velocity 
(indicated by the points) that is slightly 
damped from the beam velocity (indicated by the arrow). For 
Z$>$9 the most probable velocity of the PLF is 3.1 cm/ns, corresponding 
to a damping of $\approx$1.3 cm/ns of the beam velocity. 
This ``contamination'' of true central events with 
dissipative binary events does not significantly affect any of our subsequent
analysis. However, to provide the best isolation of true central events, 
we have additionally required in our selection of central events that 
no fragment with Z$\ge$5 is detected in the RD.

\section{Elemental yields}

\begin{figure} 
\vspace*{3.5in}
\includegraphics{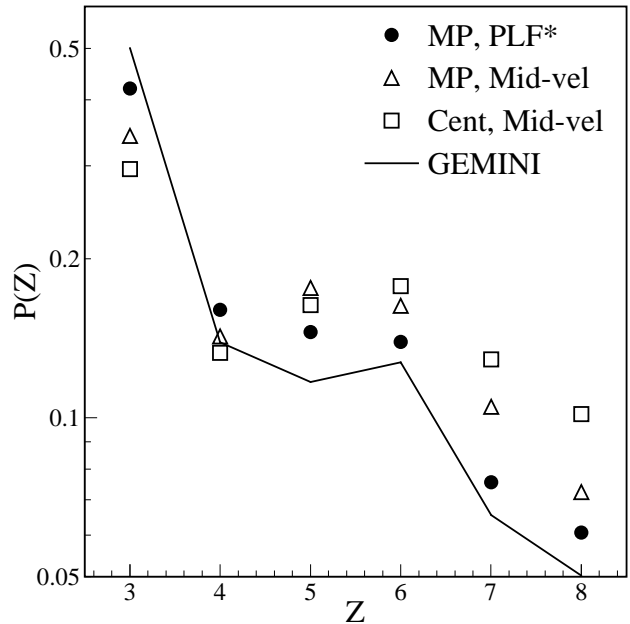} 
\caption[]
{Z distribution of fragments associated with 
mid-peripheral collisions emitted in the angular range 
85$^\circ$$\le$$\theta^{PLF}$$\le$95$^\circ$ in the PLF frame (filled circles);
mid-peripheral collisions  
and at mid-velocity (open triangles); and fragments associated with 
central collisions at mid-velocity (open squares). The lines depict the
Z distribution predicted by the statistical model GEMINI (see text for details).
The yield for each case has been normalized to unity for the interval shown.} 
\label{fig:z_distri_new}
\end{figure}

The Z distribution for IMFs associated with mid-peripheral (MP) and central (Cent) 
collisions for both PLF$^*$ emission and emission at mid-velocity is shown
in Fig.~\ref{fig:z_distri_new}. The observed yields in each case have been 
normalized for the range 3$\le$Z$\le8$.
To understand the Z distribution at mid-velocity
(0$\le$V$_{\parallel}$$\leq$1 cm/ns), we use PLF$^*$ emission (solid circles)
as a reference. Fragments emitted from the PLF$^*$ were selected 
on the basis of 
their emission angle, $\theta^{PLF}$, in the PLF frame
(85$^\circ$$\le$$\theta^{PLF}$$\le$95$^\circ$). 
For these same MP collisions, the Z distribution 
at mid-velocity (open triangles) exhibits 
a suppression of yield for Z=3 and Z=4 relative to the production of 
heavier IMFs (Z$\ge$5)
as compared to PLF$^*$ emission (solid circles).
This increase of the relative production of heavier IMFs at the expense of lighter
IMFs is even larger for the case of central collisions (open squares).
We conclude therefore that at mid-velocity
(both in mid-peripheral and central collisions), heavier fragments are produced
at the expense of lighter fragments, as compared
to PLF$^*$ emission.

This change in the Z distribution at mid-velocity 
when compared to PLF$^*$ emission, can be understood within a 
statistical framework.
Within such a framework, the Z distribution is influenced by 
the excitation, density, and size of the disintegrating system.
Statistical emission of a fragment is governed by an 
effective emission barrier relative to the temperature of the emitting 
system. Increased relative probability for the
emission of heavy fragments, can thus reflect either 
a reduction in this effective barrier and/or an increase in the 
temperature of the
system. A reduction in the density of the emitting system naturally results in
a reduction of the effective barrier. 
The Z distribution
of mid-velocity fragments in mid-peripheral collisions is
intermediate between that of PLF$^*$ emission and emission associated
with central collisions. From this observation one may conclude that within a 
statistical interpretation, the relative ``energy cost'' has already begun to change 
from that of the PLF*. 
At mid-velocity, the enhancement of yield for Z$\ge$6 in the central case 
as compared to MP collisions may be due
to the influence of finite source size.
For MP collisions,
the small size of the fragmenting system (Z$_{source}$$\approx$21) 
at mid-velocity may limit the production of heavy IMFs.

\section{Transverse velocity distributions}

\begin{figure} 
\vspace*{3.5in} 
\includegraphics{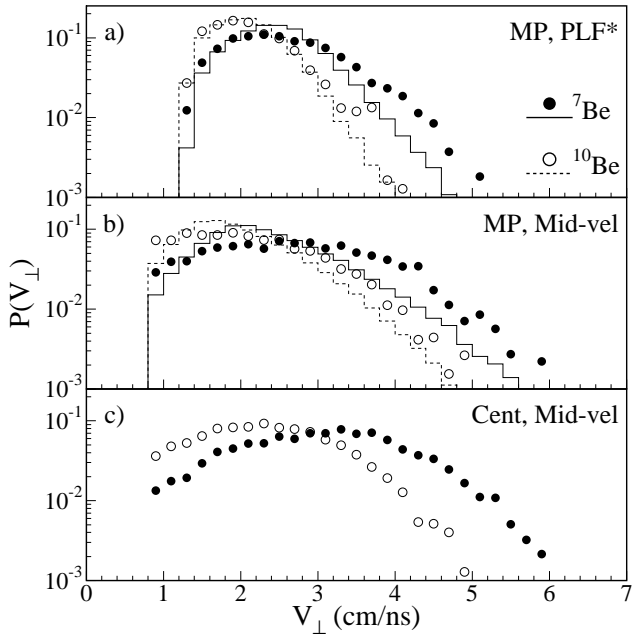}
\caption[]
{Distributions of V$_{\perp}$, for $^7$Be and $^{10}$Be fragments 
in the center-of-mass frame. 
Panel a) mid-peripheral collisions and 
V$_{\parallel}$$\geq$V$_{\parallel}$$^{PLF}$;
Panel b) mid-peripheral collisions and 0$\le$V$_{\parallel}$$\leq$1 cm/ns;
Panel c) central collisions and 0$\le$V$_{\parallel}$$\leq$1 cm/ns.
All distributions have been normalized to unity.} 
\label{fig:distri_vper_be}
\end{figure}

\begin{figure} 
\vspace*{3.5in} 
\includegraphics{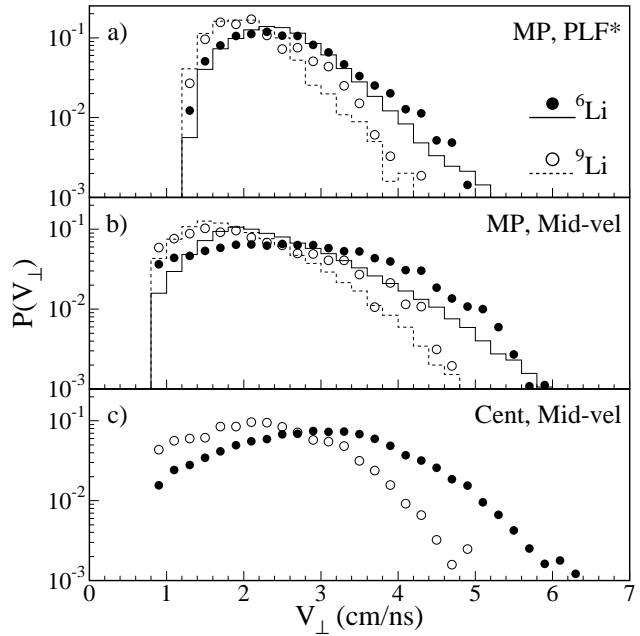}
\caption[]
{Distributions of V$_{\perp}$, for $^9$Li and $^{6}$Li fragments 
in the center-of-mass frame. 
Panel a) mid-peripheral collisions and  V$_{\parallel}$$\geq$V$_{\parallel}$$^{PLF}$;
Panel b) mid-peripheral collisions and  0$\le$V$_{\parallel}$$\leq$1 cm/ns;
Panel c) central collisions and  0$\le$V$_{\parallel}$$\leq$1 cm/ns.
All distributions have been normalized to unity.} 
\label{fig:distri_vper_li}
\end{figure}

We present in Figs.~\ref{fig:distri_vper_be}
and ~\ref{fig:distri_vper_li}
the transverse-velocity distributions of $^{7,10}$Be and $^{6,9}$Li 
fragments. The different distributions correspond to fragments 
observed in different kinematical regions for mid-peripheral
and central collisions.
Depicted in Fig.~\ref{fig:distri_vper_be}a
is the transverse-velocity distribution of $^7$Be and 
$^{10}$Be fragments 
which have parallel velocities larger than that of the PLF (in the PLF frame). 
Based on 
Fig.~\ref{fig:gali}a we understand these fragments as being  
emitted 
from the PLF$^*$. 
In this case, peaked distributions are observed 
as is expected from the
Coulomb 'ring' observed in Fig.~\ref{fig:gali}a, 
indicative of a well-defined 
Coulomb barrier characteristic of surface emission. 
The relative probability of neutron-rich $^{10}$Be 
at low V$_{\perp}$ is larger than that of
neutron-deficient $^7$Be. 
It is important to note that
the constraints of the 
experimental angular acceptance do not significantly impact the 
observed most probable velocity, an expectation confirmed by Coulomb trajectory 
calculations that will be subsequently discussed.
In contrast to the peaked distributions in Fig.~\ref{fig:distri_vper_be}a,
the distributions associated with 
mid-velocity 
emission for mid-peripheral collisions (Fig.~\ref{fig:distri_vper_be}b) 
have broader peaks and extend to larger values
of V$_{\perp}$. For V$_{\perp}$$\le$2.5 cm/ns,
the V$_{\perp}$ distribution for $^{10}$Be in particular is essentially flat. 
The observation of large yield for low V$_{\perp}$
indicates the absence of significant Coulomb repulsion in the
transverse direction. Therefore, the observed V$_{\perp}$ distributions reflect
principally the initial V$_{\perp}$ distributions. 
A broad and relatively flat
initial V$_{\perp}$ distribution is compatible
with a neck-fragmentation scenario \cite{Brosa84} or a Goldhaber picture in 
which the mid-velocity zone results from abrasion/ablation of nucleons between 
projectile and target nuclei followed by coalescence \cite{Goldhaber74,Hagel00}. 
The difference in the tails of the 
V$_{\perp}$ distribution 
between PLF$^*$ emission and mid-velocity emission
may be interpreted as the 
mid-velocity source having a higher initial temperature than the PLF$^*$ 
or possibly reflect the Fermi motion of the ablated nucleons.

Displayed in Fig.~\ref{fig:distri_vper_be}c are the V$_{\perp}$ 
distributions for fragments emitted at mid-velocity in central collisions. 
These distributions also manifest broad peaks and high 
velocity tails {\it similar} to those observed 
in Fig.~\ref{fig:distri_vper_be}b. 
Close examination however reveals that 
in this case the V$_{\perp}$ distribution is more peaked than in  
Fig.~\ref{fig:distri_vper_be}b suggesting 
that the Coulomb repulsion is larger in magnitude.
This hypothesis is qualitatively consistent with the larger
size (atomic number) of the 
source at mid-velocity for central collisions in contrast 
to mid-peripheral collisions.
On the other hand, the peak for the central case is
broader than for the case of PLF$^*$ emission shown in
Fig.~\ref{fig:distri_vper_be}a 
possibly indicating the volume breakup
of a low-density source \cite{Erin96} or surface emission from a 
hot nucleus as it expands \cite{Friedman90}.

A semi-quantitative description of the V$_{\perp}$ distributions 
presented in Fig.~\ref{fig:distri_vper_be}
can be achieved by comparing the experimental data with a N-body Coulomb 
trajectory model which simulates
the superposition of multiple source emission.
In this model, all emission was assumed to be surface 
emission originating from either the PLF$^*$ or a mid-velocity source. 
The characteristics of the PLF$^*$ were taken from the Z and energy directly measured, 
assuming the PLF$^*$ had the N/Z of the projectile. Event-by-event 
we assumed the Z of the 
mid-velocity source to be 
Z$_{mid-velocity}$ = Z$_{system}$ - Z$_{PLF}$*(1+Z$_T$/Z$_P$). We determined
$\langle$Z$_{mid-velocity}$$\rangle$= 21. Each source
also has an associated temperature that is varied independently.
The two sources were separated by an initial distance of 30 fm and 
were allowed to emit isotropically. 
Our use of
surface emission in the mid-velocity case for mid-peripheral collisions 
simply provides an ansatz for performing the 
simulation and should not be interpreted as a physical description in this 
case. This particular choice of surface emission in the 
model notably affects the 
low-velocity region of the transverse-velocity spectrum. 
While a volume 
emission scenario results in a broader distribution, the shape of the 
spectrum in the high velocity region is largely unchanged.

This simple Coulomb model
with T$_{PLF^*}$=7 MeV provides a reasonable description of
the PLF$^*$ emission as shown in Fig.~\ref{fig:distri_vper_be}a by the lines. 
Under-prediction 
of the tail of the $^7$Be distribution by this simple model is most likely due
to pre-equilibrium fragment emission processes, which are not included in the
model. As thermal energy is the only source of initial kinetic energy in the
model, reproducing the spectra depicted in Fig.~\ref{fig:distri_vper_be}b requires
T$_{mid-velocity}$=20 MeV. 
The physical origin of such a high temperature is
unclear.
The simulation also indicates that the contribution of TLF$^*$ emission
to the mid-velocity region examined is insignificant.
Analysis of the V$_{\perp}$ distributions for Li isotopes reveals similar
spectral shapes and slope parameters 
(T$_{PLF^*}$=7 MeV and T$_{mid-velocity}$=20 MeV)
as shown in Fig. ~\ref{fig:distri_vper_li}. 
Kinetic ``temperatures'' of similar magnitude 
have been previously
reported for mid-velocity IMF emissions \cite{Piantelli02,Lukasik02}. 

It is noteworthy that both the
$^7$Be and $^{10}$Be, as well as the $^{6,9}$Li, spectra 
associated with mid-velocity emission, shown in Figs.   ~\ref{fig:distri_vper_be} and ~\ref{fig:distri_vper_li},  
can be described with the same initial 
kinetic energy (T$_{mid-velocity}$=20 MeV).
The enhanced 
probability of low V$_\perp$ emission as compared to the 
surface emission model 
may reflect temperature-dependent surface-entropy effects
\cite{Toke03}  or a 
transition from surface to volume emission \cite{Beaulieu00}.

\section{Isotopic composition}

\begin{figure} 
\vspace*{3.5in} 
\includegraphics{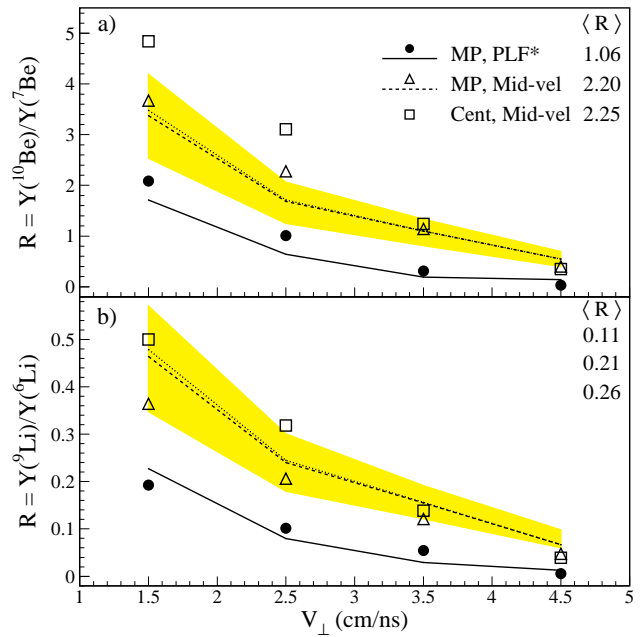}
\caption[]
{Dependence of the ratios of $^{10}$Be to $^7$Be and $^{9}$Li to $^6$Li on V$_{\perp}$ for different impact parameter and velocity selection criteria. 
Points displayed indicate the average value for
1$\le$V$_\perp$$<$2, 2$\le$V$_\perp$$<$3, 3$\le$V$_\perp$$<$4, 
and V$_\perp$$\ge$4 cm/ns. 
Lines correspond to the 
predictions of a two source emission model. The shaded region corresponds to 
different assumptions for the neutron-enrichment of mid-velocity.
} 
\label{fig:iso_vt_be}
\end{figure}

The experimental observation that the transverse-velocity distributions for the 
neutron-rich (e.g. $^{10}$Be and $^9$Li) isotopes are different from the 
neutron-deficient isotopes (e.g. $^{7}$Be and $^6$Li) 
(Figs. ~\ref{fig:distri_vper_be} and
~\ref{fig:distri_vper_li})
indicates that the ratio of these isotopes evolves as a function of V$_{\perp}$.
In order to examine this dependence 
in a more transparent fashion, 
we examine the ratio of the yields of $^{10}$Be to $^7$Be 
as a function of V$_\perp$ in Fig.~\ref{fig:iso_vt_be}a. 
For all three cases shown one observes that this ratio 
decreases monotonically with increasing V$_{\perp}$. 
For 
the case of PLF$^*$ emission (solid circles) this decrease  
can be simply understood as a consequence of the Coulomb barrier. 
As the Coulomb barrier for both $^{10}$Be and $^7$Be emission is similar, the
$^7$Be, on average, acquires a higher velocity than the $^{10}$Be. 
The region of low V$_\perp$ is therefore preferentially populated by the 
$^{10}$Be as compared to the $^7$Be as evident in 
Fig. ~\ref{fig:distri_vper_be}a.

The most striking feature of Fig.~\ref{fig:iso_vt_be}a is that 
mid-velocity fragments associated with mid-peripheral collisions 
(open triangles) exhibit a significantly larger value of 
Y($^{10}$Be)/Y($^7$Be) as compared to emission from the PLF$^*$ (solid circles). 
This enhancement is observed at all values
of V$_{\perp}$. 
Mid-velocity fragments associated with central collisions (open squares)
also manifest large 
values of Y($^{10}$Be)/Y($^7$Be), as compared to emission from the PLF$^*$.
The yield ratios associated with central collisions are even larger than 
those mid-velocity emission in the  
mid-peripheral case (open triangles), however 
most of this difference occurs for low V$_{\perp}$ (V$_{\perp}$ $\le$2.5 cm/ns).
This V$_{\perp}$ dependence of  Y($^{10}$Be)/Y($^7$Be) for both PLF$^*$ 
emission
and mid-velocity emission can explain
the angular dependence of neutron-deficient fragments previously
reported \cite{Ramakrishnan98}. As different Coulomb repulsion in the 
different cases leads to different behavior of the yield ratio as a function
of V$_{\perp}$, it is important to compare the yield ratios integrated over 
V$_{\perp}$.

By integrating over the entire
range of V$_{\perp}$ observed, we find that $\langle$Y($^{10}$Be)/Y($^7$Be)$\rangle$ for 
PLF$^*$ emission is 1.06 while at mid-rapidity it is 
considerably higher, 2.2 for
mid-peripheral collisions and 2.25 for central collisions. Based upon these
integrated yields, one observes that for Be,
\begin{equation}
\frac{R_{mid-velocity,mid-peripheral}}{R_{PLF,mid-peripheral}}=
\frac{2.2}{1.06} = 2.08
\end{equation}
while  
\begin{equation}
\frac{R_{mid-velocity,central}}{R_{PLF,mid-peripheral}}=
\frac{2.25}{1.06} = 2.12
\end{equation}
{ \it Thus, the integrated yield of} $^{10}$Be/$^7$Be 
{\it at mid-velocity for mid-peripheral and central collisions is 
comparable and significantly different from PLF$^*$ emission.}
The behavior exhibited by the Y($^{10}$Be)/Y($^7$Be) is also observed
for Y($^{9}$Li)/Y($^6$Li), as shown in the lower panel of Fig.
~\ref{fig:iso_vt_be}.
{\it It is remarkable that at mid-velocity not only are the Z and transverse-velocity distributions similar for mid-peripheral and central collisions, but the fragment composition is essentially the same -- while markedly different from PLF$^*$ emission.}

It is important to consider the influence of the PLF$^*$ Coulomb field on 
the observed N/Z enrichment at mid-velocity.
Radial repulsion of fragments from the PLF$^*$ results in $^7$Be 
fragments displaced more toward mid-velocity than 
$^{10}$Be fragments. Therefore the Coulomb field of the PLF$^*$ and TLF$^*$
lead to an increase in the
yield of neutron-deficient isotopes at mid-velocity. 
Hence, due to these qualitative arguments 
the primordial N/Z composition at mid-velocity is
expected to be higher than that experimentally observed.

The N-body Coulomb trajectory model previously described can also be used 
to quantify this enhancement of 
Y($^{10}$Be)/Y($^{7}$Be) at mid-velocity as compared to PLF$^*$ emission.
The relative emission probability, Y($^{10}$Be)/Y($^7$Be), 
for the PLF$^*$ was taken from the
experimental data while for the mid-velocity source this probability 
was taken relative to the PLF$^*$ ratio 
as K*Y$_{PLF^*}$($^{10}$Be)/Y$_{PLF^*}$($^7$Be), where K
was varied as a free parameter. 
For mid-peripheral collisions, the {\it superposition} of 
PLF$^*$ emission and emission
of the mid-velocity source with V$_\parallel$$>$V$_\parallel$$^{PLF}$ 
in the model is 
indicated by the solid line in Fig.~\ref{fig:iso_vt_be}, 
while the corresponding yield at 
mid-velocity is represented by the dashed line. 
Reproducing the V$_{\perp}$ dependence of the relative
yield, requires consideration of both the Coulomb repulsion from the emitting source, 
and the initial 
velocities of the fragments.
Accounting for the initial velocities in a realistic manner is accomplished 
by attributing a temperature to the emitting source.
In the case of emission from the PLF$^*$ 
we use, consistent with the 
V$_\perp$ of Fig.~\ref{fig:distri_vper_be}a, a temperature of
T$_{PLF^*}$=7 MeV to model the PLF$^*$ decay.
The solid line presented in Fig.~\ref{fig:iso_vt_be}
reflects 
predominantly the emission from the PLF$^*$. 
Reproducing the mid-velocity data necessitates that  
Y$_{mid-velocity}$($^{10}$Be/$^{7}$Be)={\bf 2}*
Y$_{PLF^*}$($^{10}$Be/$^7$Be)
{\it and} that T$_{mid-velocity}$=20 MeV. 
These latter temperatures are also consistent with those deduced from the 
V$_{\perp}$ distributions shown in Figs.~\ref{fig:distri_vper_be} and
~\ref{fig:distri_vper_li}.
Shown as the shaded region in Fig.~\ref{fig:iso_vt_be} is the 
sensitivity of the yield ratio to the parameter K. The bottom of the shaded 
region corresponds to K=1.5 while the top corresponds to K=2.5.
While the dashed line represents the superposition of mid-velocity {\it and}
PLF$^*$ emission, the dotted line depicts the contribution of only the 
mid-velocity source.
While the dotted line is only slightly higher than the
dashed line, this difference is sensitive to the assumption of isotropic
emission from the PLF$^*$. Enhanced
backward emission from the PLF$^*$, as has recently been 
reported \cite{Hudan04}, would correspondingly result in a larger 
difference. 
All the trends described for  
Y($^{10}$Be)/Y($^7$Be) are also observed for Y($^{9}$Li)/Y($^6$Li),
as shown in Fig. ~\ref{fig:iso_vt_be}b, supporting
the conclusion that the Y($^{10}$Be)/Y($^7$Be) ratio is not an isolated case
but is representative of other fragments. 
To extract more quantitative information on the relative yield enhancement 
at mid-velocity, it is necessary to more accurately account 
for the ``distortion'' 
introduced by the presence of  
the PLF$^*$ and TLF$^*$. 
The distortion, particularly important for low V$_{\perp}$, 
depends on the IMF emission rates 
relative to the PLF$^*$-TLF$^*$ re-separation and is beyond the scope of the
present analysis.

\section{Gemini calculations}

To understand the PLF$^*$ decay better,  
we performed calculations with the statistical-model code GEMINI 
\cite{Charity01}, which describes 
surface emission from an excited nucleus including emission from 
excited states and their sequential decay. We examined the
isotopic yields as a function of the excitation and spin
of a parent nucleus. At each 
excitation energy we roughly reproduce the average atomic number of the
measured PLF. 
In Table I, for a fixed J one observes that the 
Y($^{10}$Be)/Y($^{7}$Be) decreases with increasing E*/A,
while for fixed E*/A, it increases with increasing J. For J = 0$\hbar$ to 
reproduce the measured value of 1.06 associated with PLF$^*$ emission,
we deduce an excitation of E*/A $\approx$ 3-4 MeV. As the level density is 
taken to be 
a=A/9 MeV$^{-1}$ in the model, this excitation corresponds to a temperature of 
T$\approx$6 MeV, in 
reasonable agreement with the tails of the transverse-velocity distribution. 
If the PLF$^*$ has significant spin 
($\approx$20$\hbar$), the temperature of the source must be somewhat higher.
Thus, within a surface emission picture the Y($^{10}$Be)/Y($^7$Be) and the 
transverse-velocity distributions
constrain the excitation and spin of the emitting source. Correctly 
accounting for the interplay of both of these quantities is necessary 
to describe the conditions of the emitting source.
For these conditions (Z=48, A=114, E$^*$/A=3.5 MeV, J=0$\hbar$), 
we have compared the yield distribution of fragments predicted by GEMINI with that 
observed for PLF$^*$ emission. The solid line shown in 
Fig.~\ref{fig:z_distri_new} is the GEMINI yield distribution normalized to 
the range 3$\le$Z$\le$8.
While in general the agreement between the GEMINI calculations and emission from the
PLF$^*$ is reasonable, GEMINI overpredicts the relative yield of Z=3 while 
under-predicting  the relative yield for Z$\ge$4. An improved description of 
PLF$^*$ emission could be done with a
meta-stable mononuclear model which accounts for its expansion \cite{Sobotka04}.
While this may affect the overall emission probabilities, such a treatment 
is unlikely to affect the ratios presently discussed.

\begin{table}
\begin{ruledtabular}

\begin{tabular}{ccccc}
E$^*$/A(MeV)&J=0$\hbar$&J=20$\hbar$&J=40$\hbar$&
\\
\hline
2& 1.50 & $>$1.8 & 4.67 \\
3& 1.29 & 1.14 & 2.11 \\
4& 0.78 & 1.17 & 1.42 \\
\end{tabular}
\end{ruledtabular}
\caption{Results of GEMINI calculations indicating the 
dependence of Y($^{10}$Be)/Y($^7$Be) on excitation energy, 
E$^*$/A, and spin, J, for emission 
from the PLF$^*$ (Z$\approx$ 40-48, N/Z $\approx$ 1.375).}
\end{table}

\section{Z and A dependence of $\langle$E$_\perp$$\rangle$}

\begin{figure} 
\vspace*{3.5in} 
\includegraphics{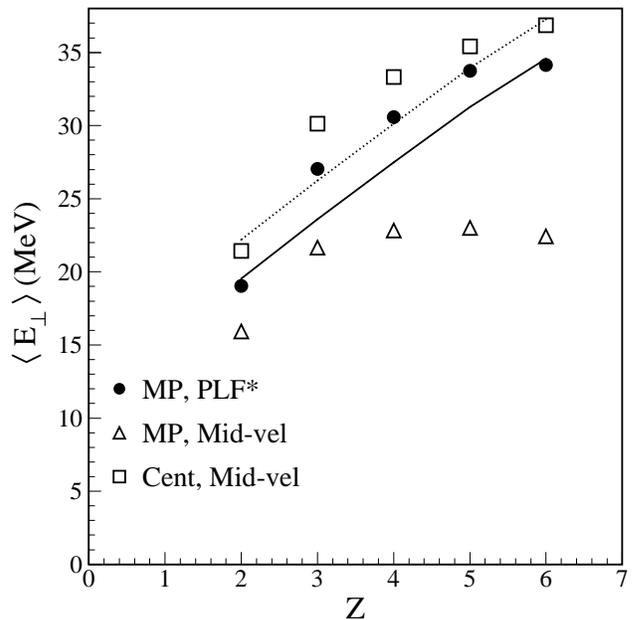}
\caption[]
{Average transverse energies for 2$\le$Z$\le$6 selected under different criteria.
For the PLF$^*$, fragments were selected in the angular range 
85$^\circ$$\le$$\theta^{PLF}$$\le$95$^\circ$ while mid-velocity 
fragments have 0$\le$V$_{\parallel}$$\le$1cm/ns in the center-of-mass frame. 
}
\label{fig:average_et}
\end{figure}

The average transverse energies of 2$\le$Z$\le$6 fragments are displayed 
in  Fig.~\ref{fig:average_et}
for different impact parameter and velocity selection criteria.
To examine emission from the PLF$^*$ without being biased by the 
minimum angular acceptance of LASSA, we have selected emission for
85$^\circ$$\le\theta^{PLF}\le$95$^\circ$ in the PLF frame.
The values of $\langle$E$_{\perp}$$\rangle$ for these different selections
can be understood in the context of a simple physical picture 
in which fragments 
acquire their transverse kinetic energy as a result of Coulomb repulsion, 
thermal motion, and possibly energy associated with collective expansion 
\cite{deSouza93}. 
As the thermal component is mass (Z) independent, an observed Z dependence 
can be attributed to Coulomb repulsion of fragments from the emitting PLF$^*$
or the presence of collective motion \cite{deSouza93}.
Distinguishing between the Z and A dependence of the $\langle$E$_{\perp}$$\rangle$ 
can differentiate between the Coulomb and collective contributions to the 
transverse energy.
In the case of emission from the PLF$^*$ (solid circles), the measured
average transverse kinetic energies in the PLF frame generally increase with Z.
The magnitude of $\langle$E$_{\perp}$$\rangle$ for Z=2 is roughly consistent with
values previously reported for a similar system \cite{Wieloch98}.
These measured magnitudes of $\langle$E$_{\perp}$$\rangle$  can be understood
within a surface emission picture in which fragments are emitted from the 
surface of a hot nucleus. The $\langle$E$_{\perp}$$\rangle$ in such a picture is given by
$\langle$E$_{CM}$$\rangle$ = $\langle$V$_{C}$$\rangle$ + 2 $\langle$T$\rangle$ where
the Coulomb energy V$_C$= 1.44 (Z$_{source}$-Z$_{fragment}$)*Z$_{fragment}$/R
with R, the center-to-center separation distance between the fragment and 
source, taken as 
R=1.2*((A$_{source}$-A$_{fragment}$)$^{1/3}$+A$_{fragment}$$^{1/3}$) + 2 fm.
The average Coulomb energy has been calculated within this simple picture.       The corresponding $\langle$E$_{\perp}$$\rangle$  
($\langle$E$_{\perp}$$\rangle$ =  $\frac{2}{3}$$\langle$E$_{CM}$$\rangle$)
for
a temperature of T=7 MeV is indicated by the solid line in 
Fig.~\ref{fig:average_et} while T=9 MeV is represented by the dotted line. 
The magnitude of the observed $\langle$E$_{\perp}$$\rangle$ and the dependence on Z 
is qualitatively described by this simple model, 
bolstering the view that this emission
is standard statistical emission from a near normal density PLF$^*$.

In Fig. ~\ref{fig:average_et}, a
marked contrast between emission from the PLF$^*$ and fragments at mid-velocity 
associated with mid-peripheral 
collisions (open triangles) is observed.
Mid-velocity fragments in MP collisions 
manifest significantly lower $\langle$E$_{\perp}$$\rangle$
in the PLF frame and for 3$\le$Z$\le$6 
essentially no dependence of $\langle$E$_{\perp}$$\rangle$ on Z exists.
The $\langle$E$_{\perp}$$\rangle$ for helium is approximately 25-30$\%$ lower
than for IMFs.
The magnitude of $\langle$E$_{\perp}$$\rangle$ for 3$\le$Z$\le$6 is 
$\approx$22 MeV, essentially  2/3 of the Fermi energy.  
The magnitude of $\langle$E$_{\perp}$$\rangle$ for Z=2 is $\approx$16 MeV.
$\langle$E$_{\perp}$$\rangle$ values of similar magnitude at mid-velocity 
have previously been 
reported for IMFs, integrated over Z, and neutron-rich light-charged-particles
for a significantly lighter system \cite{Lefort00}.
This independence on Z, together with the 
fact that the $\langle$E$_{\perp}$$\rangle$ is approximately 2/3 the Fermi energy 
suggests 
that the transverse energy of these fragments does not contain any significant 
Coulomb contribution, as was previously conjectured based upon the
V$_{\perp}$ distribution shown in Figs.~\ref{fig:distri_vper_be} and
~\ref{fig:distri_vper_li}.  This result is consistent with a physical picture 
in which fragments aggregate in a dilute nuclear medium -- 
compatible with a Goldhaber scenario as has been previously 
suggested \cite{Lukasik02}. Both the Z independence of 
$\langle$E$_{\perp}$$\rangle$ and the reported magnitudes in the present work 
are consistent with those previously reported \cite{Lukasik02}.
Alternatively, this behavior can be viewed as
the volume breakup of a 
low-density source or as emission from a distended configuration.
On the
basis of transverse energies, it has been proposed 
\cite{Piantelli02,Lukasik02} that 
fragments intermediate between the PLF$^*$ and TLF$^*$ are 
dynamical in nature, maintaining early-stage correlations of the 
collision \cite{Nebauer99}. 

For mid-velocity fragments associated with central collisions (open squares),
examination of the dependence of $\langle$E$_{\perp}$$\rangle$ on Z, 
and in particular the magnitudes of the measured  
$\langle$E$_{\perp}$$\rangle$ is particularly revealing.
For this case, $\langle$E$_{\perp}$$\rangle$ increases monotonically with Z, indicating
a Coulomb influence. For Z=2, the value measured in this work is consistent
with values previously reported \cite{Wieloch98}. The magnitude of 
$\langle$E$_{\perp}$$\rangle$, however, is only slightly larger than 
in the case of emission from the PLF$^*$ (solid circles). 
These two cases, however, involve significantly different charge associated with the
emitting system.
For central collisions
the initial atomic number of the mid-velocity source 
is almost double that of the PLF$^*$ (Z$_S$=72 as compared to 
Z$_{PLF*}$=41). Hence, the similarity of the values of
$\langle$E$_{\perp}$$\rangle$ for central collisions with those for
emission from the PLF$^*$ suggests that
for central collisions, fragments
originate either from a dilute nuclear system (either volume breakup or surface 
emission during expansion), or after considerable 
charge has been removed from the system, by fast, light-charged-particle 
emission. 
The large Coulomb barrier for IMF emission favors early emission of IMFs
making IMF emission following 
light-charged-particle de-excitation, on average, less likely.

\begin{figure}
\vspace*{3.5in} 
\includegraphics{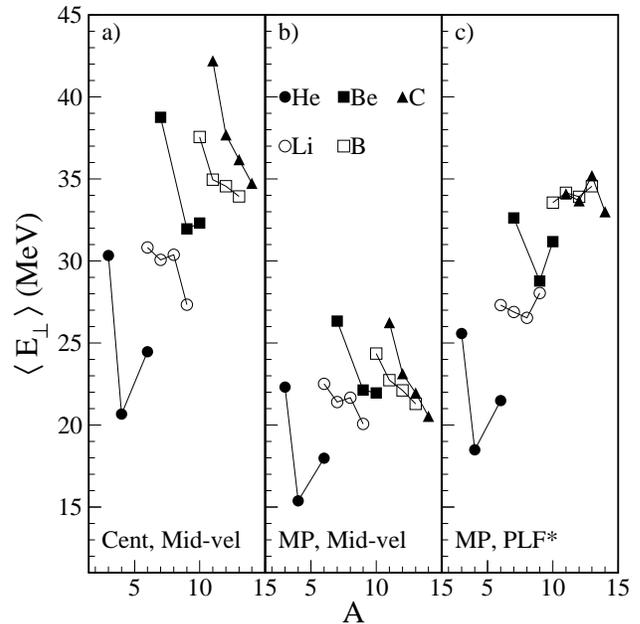}
\caption[]
{Average transverse energies as a function of mass number for 2$\le$Z$\le$6 
for mid-peripheral and central collisions with different selection criteria.
For the PLF$^*$, fragments were selected in the angular range 
85$^\circ$$\le$$\theta^{PLF}$$\le$95$^\circ$ while mid-velocity fragments
have 0$\le$V$_{\parallel}$$\le$1cm/ns in the center-of-mass frame. 
} 
\label{fig:ave_et_mass}
\end{figure}

To disentangle the contribution of Coulomb energy and any possible collective
flow \cite{deSouza93} to the $\langle$E$_{\perp}$$\rangle$, 
we examine the dependence of $\langle$E$_{\perp}$$\rangle$ on A 
for individual elements in
Fig.~\ref{fig:ave_et_mass}. 
Significant collective expansion effects have been previously 
asserted \cite{Pak96,Marie97}, 
particularly for the case of central collisions.
Evident in panel a) for the case of 
central collisions is that for IMFs, in the case of N$\ge$Z, the observed 
$\langle$E$_{\perp}$$\rangle$ does not increase with increasing A for a given
Z but is either constant or decreases slightly. This observation contradicts
the expectation of a mass-dependent collective flow, 
at least in the transverse direction.
The most striking feature of 
Fig.~\ref{fig:ave_et_mass}a is that
neutron-deficient isotopes, particularly $^3$He, $^7$Be, and $^{11}$C, exhibit 
larger $\langle$E$_{\perp}$$\rangle$ than other isotopes of the same element. 
For $^6$Li essentially no enhancement is observed while for $^{10}$B only a 
modest enhancement is observed. 
One possible reason for this difference between odd and even Z is that 
only for even-Z are nuclides with N$<$Z observed  
with significant yield. 
This enhancement in the kinetic energy of 
neutron-deficient fragments in comparison to other isotopes of the same element
has been previously reported for central collisions in the 
system $^{112}$Sn + $^{112}$Sn at
E/A=50 MeV and has been interpreted as evidence for surface emission from a
hot, expanding nuclear system \cite{Liu04}. 
It has been hypothesized that the larger 
kinetic energy observed for the neutron-deficient isotopes originates 
because they are emitted on average earlier than other isotopes of the 
same element \cite{Liu04}. 

In contrast to the previous work 
which focused solely on central collisions \cite{Liu04}, 
we also present the dependence of $\langle$E$_{\perp}$$\rangle$ 
associated with mid-peripheral collisions for both 
emission from the
PLF$^*$ and the mid-velocity regime. In the case of mid-velocity 
we observe the same effect as for central collisions, although the 
magnitude of the enhancement is somewhat less. 
As the physical picture of a hot, source which emits as it expands is 
not compatible with the case of mid-velocity emission for mid-peripheral 
collisions, the 
observed trend must have an alternate explanation.
In the case of 
emission from the PLF$^*$,
$^{11}$C does {\it not} show an enhancement in contrast to the two
mid-velocity cases, emphasizing the difference between mid-velocity fragment
production and emission from the PLF$^*$.
A significant enhancement of the average kinetic energy is 
still observed
for $^3$He and $^7$Be emitted from the PLF$^*$. 
Thus, this kinetic enhancement of neutron-deficient isotopes
is not associated simply with central collisions in which a low density, expanded
source is formed but also occurs in the standard statistical decay of a near 
normal density PLF$^*$.

\begin{figure}
\vspace*{3.5in} 
\includegraphics{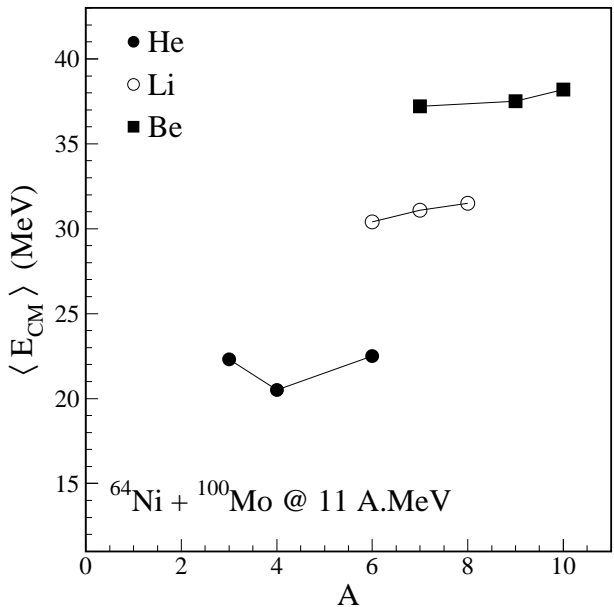}
\caption[]
{
Average center-of-mass energies for 2$\le$Z$\le$4 as a function of mass 
selected by element following incomplete fusion in the reaction $^{64}$Ni +
$^{100}$Mo at E/A=11 MeV \cite{Charity01}.} 
 
\label{fig:ave_ecm_mass_bob}
\end{figure}

If the observed effect is related to 
a displacement in the
emission time distributions of neutron-deficient isotopes with respect to 
heavier isotopes of the same element \cite{Liu04}, 
one would expect the effect to decrease with
increasing excitation energy
as the system moves toward instantaneous 
breakup \cite{Beaulieu00}. 
As a baseline for a nuclear system at relatively low excitation, 
we have examined helium isotopes and
IMFs emitted in the reaction $^{64}$Ni +$^{100}$Mo at E/A=11 MeV 
\cite{Charity01}.
Following incomplete fusion of the projectile Ni nuclei with the Mo target 
nuclei, evaporation residues were measured in coincidence with emitted
neutrons, charged particles, and IMFs detected at selected angles. 
The kinetic energy spectra of 
these emitted particles is clearly evaporative. 
On the basis of its velocity, the excitation of the 
evaporation residue is estimated 
to be E$^*$=319$\pm$27 MeV \cite{Charity04}. Thus, this system  
provides an important low-excitation reference point in a system of comparable
Z and A to the central collision data of the present work.
The dependence of $\langle$E$_{CM}$$\rangle$ on A is shown in 
Fig.~\ref{fig:ave_ecm_mass_bob}. Clearly evident in 
this figure is the fact that 
standard 
statistical emission at low excitation {\it does not} result in a large 
enhancement of the kinetic energies of the neutron-deficient 
isotopes $^3$He and $^7$Be.
A difference of $\approx$2 MeV is observed between the $^3$He 
and $^4$He average kinetic 
energies. Due to the low excitation energy of this system, 
sequential feeding of $^3$He is
expected to be negligible. Consequently, the observed energy 
difference between $^3$He and $^4$He can be 
largely attributed to the earlier average emission time of $^3$He.
This difference provides a reference point 
for the increase in cluster kinetic energy 
due to differences
in the average emission time. 
The large difference observed in 
average kinetic energies of neutron-deficient isotopes in
the $^{114}$Cd + $^{92}$Mo system 
is therefore not principally due 
to differences in the average emission time of fragments.

\begin{figure}
\vspace*{3.5in} 
\includegraphics{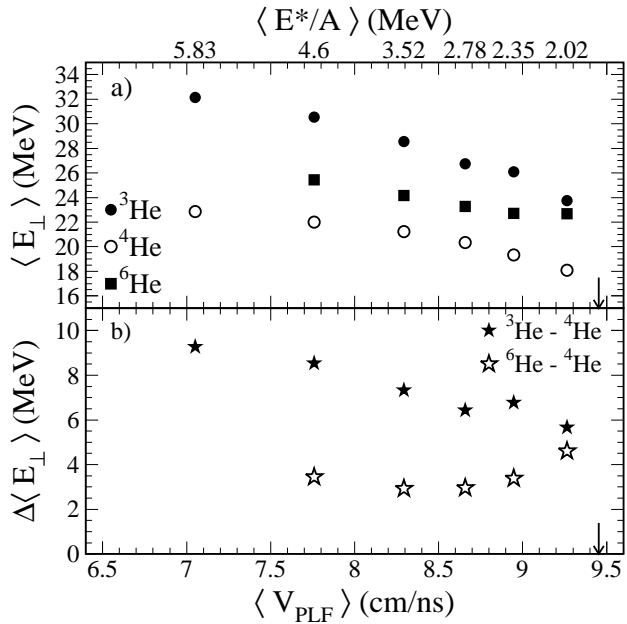}
\caption[]
{Panel a) $\langle$E$_{\perp}$$\rangle$ for helium isotopes as a function of
$\langle$V$_{PLF}$$\rangle$ and deduced excitation energy (upper axis). 
$^3$He (solid circles), $^4$He (open circles), and $^6$He (solid squares). 
Panel b) Average transverse energy of $^3$He 
and $^6$He with reference to $^4$He.} 
\label{fig:he_damping}
\end{figure}

Direct evidence that the excitation of the emitting source is 
primarily responsible 
for the enhancement of the neutron deficient isotopes is presented in 
Fig.~\ref{fig:he_damping} for emission from the PLF$^*$. 
In panel a) the $\langle$E$_{\perp}$$\rangle$ 
for isotopes of
helium are shown as a function of the PLF velocity. 
For this portion of the analysis we expand our definition of the PLF to include
15$\le$Z$_{PLF}$$\le$46 \cite{Yanez03}.
It has previously 
been demonstrated that the velocity damping of 
the PLF$^*$ is associated with the excitation incurred in the PLF$^*$ 
following the interaction phase of the collision \cite{Yanez03}. 
The deduced excitation energy scale  
is indicated at the top of the figure while 
the beam velocity is represented by an arrow. As 
$\langle$V$_{PLF}$$\rangle$ decreases from the beam velocity 
(excitation energy increases)
the $\langle$E$_{\perp}$$\rangle$ for
$^3$He, $^4$He, and $^6$He all increase monotonically. 
To explore the differences between the increase in kinetic energy for
the different helium isotopes in a more sensitive manner, we examined
the increase in the $\langle$E$_{\perp}$$\rangle$ for 
$^3$He and $^6$He relative to $^4$He. 
These results are presented in the lower panel of 
Fig.~\ref{fig:he_damping}. The difference in transverse energy between
$^3$He and $^4$He increases with increasing excitation from 5.7 MeV at 
$\langle$E*/A$\rangle$$\approx$2 MeV to 9.3 MeV at $\langle$E*/A$\rangle$$\approx$5.8 MeV. On the 
other hand, aside from an initial decrease, the transverse energy 
difference between 
$^6$He and $^4$He relative remains approximately constant over the 
measured range.
We emphasize that the 
average kinetic energy enhancement for
neutron-deficient isotopes increases with increasing 
excitation energy (solid stars),
opposing the 
expected behavior based upon an emission time displacement argument.

These observations may be qualitatively understood 
by considering the growing importance of 
charged-particle decay channels with increasing excitation energy. A fragment
emitted from a hot source with an initial thermal energy accelerates in the
Coulomb field of the emitting system and acquires its asymptotic 
kinetic energy. If this fragment decays by neutron emission, 
the velocity of the secondary fragment is
on average the same as the primary fragment
at the moment of decay, 
thus its observed kinetic energy is only impacted by the change in mass.
However, should the fragment undergo charged-particle decay, then the 
kinetic energy observed for the secondary fragment reflects the Coulomb 
energy acquired by its parent up to the moment of decay, 
which is larger due to the  
larger parent atomic number. {\it Only if} the lifetime of the 
parent is sufficiently long for it to transform a significant fraction 
of its initial Coulomb energy into kinetic energy will the kinetic energy of the
daughter fragment be appreciably increased. Instantaneous decay
of the parent fragment will not result in an increase in the 
kinetic energy of the
daughter fragments. This physical picture suggests that the neutron-deficient
isotopes manifest a secondary decay contribution from relatively long-lived
charged-particle channels, i.e narrow resonances at relatively low 
excitation in the parent fragment. 
It is therefore 
clear from the evidence presented that the fragments are not created 
relatively cold
as predicted in some multi-fragmentation models \cite{Gross90}. Moreover, these
hot fragments do not decay instantaneously. Measurement of the yield 
associated with multi-particle resonant decay would provide 
quantitative information about this scenario. Unfortunately,
the present experimental data does not allow examination of these 
resonant decays. 
With increasing excitation, the secondary decay feeding to
$^3$He and $^4$He changes, presumably affects the yield ratios of
the two isotopes a reference benchmark for isotope thermometry
\cite{Albergo85,Pochodzalla95}. It should be realized that independent of 
the underlying origin of the kinetic energy difference,
there is an 
inherent danger
of examining double ratios such as (Y($^3$He)/Y($^4$He))/(Y($^6$Li)/Y($^7$Li))
involving only one isotope with N$<$Z that manifests a considerably
different kinetic energy.
Our results may also suggest that 
extracting the primordial N/Z to investigate the possible isospin 
fractionation of a dilute nuclear medium requires detailed measurement of
both neutron and charged-particle resonant decays.

\section{Conclusions}

In summary, it is revealing to compare the characteristics of
mid-velocity fragments and those emitted in the de-excitation of a 
hot, near-normal density nucleus, namely the PLF$^*$.
The Z distributions and the transverse-velocity distributions,
for mid-velocity emission are different from those associated with 
PLF$^*$ emission. 
On the
other hand, fragments observed at mid-velocity are rather similar independent of
whether they are associated with mid-peripheral or central collisions. 
The integrated yield ratios of $^{10}$Be/$^{7}$Be and $^{9}$Li/$^{6}$Li
reveal that mid-velocity and PLF$^*$ emission are 
also substantially different in N/Z. 
For central collisions and mid-peripheral collisions, 
these observed yield ratios
are enhanced by a factor of approximately two with respect to PLF$^*$ emission.
In the case of emission from the PLF$^*$, the Z dependence of 
$\langle$E$_\perp$$\rangle$ shows that  
fragment emission is {\it consistent} with standard evaporation from near 
normal density nuclear matter. 
In central collisions the Z dependence and magnitude of
$\langle$E$_\perp$$\rangle$ for mid-velocity fragments
are {\it inconsistent} with normal density formation. 
For mid-velocity
fragments formed in mid-peripheral collisions, the Z independence of
$\langle$E$_\perp$$\rangle$ and a magnitude of approximately two-thirds of the
Fermi energy, suggest cluster formation through coalescence of ablated 
nucleons.
All these facts are consistent with the low-density formation of fragments at
mid-velocity for both mid-peripheral and central collisions, indicating that 
the conditions for fragment formation at mid-velocity are 
{\it significantly different from those of PLF$^*$ emission.} 
Examination of $\langle$E$_\perp$$\rangle$ for 
isotopically identified
fragments shows that neutron-deficient isotopes, particularly those with 
N$<$Z, exhibit larger kinetic energies than heavier isotopes of the 
same element. 
This enhancement of the kinetic energies for neutron-deficient isotopes
increases with increasing excitation 
energy. This result suggests that 
fragments are produced hot and that long-lived
charged-particle decay may be important for N$<$Z clusters, motivating the
future study of resonant decay.

\begin{acknowledgments}

	We would like to acknowledge the 
valuable assistance of the staff at MSU-NSCL for
providing the high quality beams which made this experiment possible. 
This work was supported by the
U.S. Department of Energy under DE-FG02-92ER40714 (IU), 
DE-FG02-87ER-40316 (WU) and the
National Science Foundation under Grant No. PHY-95-28844 (MSU).\par
\end{acknowledgments}

\bibliography{hudan5.bib} 

\end{document}